\documentclass[showpacs,aps,preprint]{revtex4}
\usepackage{graphicx}
\usepackage{amsmath,amssymb}

\usepackage{color}

\def\gtsim{\mathrel{\hbox{\raise0.2ex
  \hbox{$>$}\kern-0.75em\raise-0.9ex\hbox{$\sim$}}}}
\def\ltsim{\mathrel{\hbox{\raise0.2ex
  \hbox{$<$}\kern-0.75em\raise-0.9ex\hbox{$\sim$}}}}

\def\mbold#1{\mbox{\boldmath $#1$}}

\newcommand{\rs}[1]{_{\rm #1}} 
\newcommand{\bm}[1]{\mbox{\boldmath $#1$}} 

\begin{document}

\title{
Microscopic study of tetrahedrally symmetric nuclei by an angular-momentum and parity projection method}

\author{Shingo Tagami and Yoshifumi R. Shimizu}
\affiliation{Department of Physics, Faculty of Sciences,
Kyushu University, Fukuoka 812-8581, Japan}
\author{Jerzy Dudek}
\affiliation{
Institut Pluridisciplinaire Hubert Curien (IPHC),
IN$_2$P$_3$-CNRS/Universit\'e de Strasbourg, F-67037 Strasbourg, France }

\begin{abstract}

We study the properties of the nuclear rotational excitations
with hypothetical tetrahedral symmetry
by employing the microscopic mean-field and residual-interaction Hamiltonians
with angular-momentum and parity projection method;
we focus on the deformed nuclei with tetrahedral
doubly-closed shell configurations.
We find that for pure tetrahedral deformation
the obtained excitation patterns satisfy the characteristic features
predicted by group-representation theory applied to
the tetrahedral symmetry group.  We find that a gradual transition
from the approximately linear to the characteristic rigid-rotor,
parabolic energy-vs.-spin dependence occurs
as a function of the tetrahedral deformation parameter.
The form of this transition is compared
with the similar well-known transition in the case of quadrupole deformation.

\end{abstract}


\maketitle


\section{Introduction}
\label{sec:intro}

Symmetries play an important role in physics, often serving as guide-lines
in studying characteristic features of motion of quantum systems.
In particular, in nuclear physics, the spatial symmetries of the nuclear
mean-field potential are crucial in determining both the properties of
the independent particle motion and of the nuclear stability.
The best known, the spherical symmetry of the potential, implies
the high degeneracy of the single-particle level energies,
the so-called magnetic degeneracy, $(2j+1)$, of the orbitals characterised
by the spherical-shell angular-momentum quantum-number, $j$,
and leads to the well-known spherical magic numbers in nuclei.
The spontaneous breaking of the spherical symmetry arises
when the energies of the spherical configuration
of the system with certain particle numbers are higher as compared
to the energies of the alternative non-spherical spatial distribution
of nucleons, the mechanism to which the Pauli exclusion principle
contributes importantly. The spherical symmetry breaking removes
the $(2j+1)$-degeneracy and leads to the deformed single-particle
orbital scheme~\cite{BM75,RS80}.

More generally, nuclei which are not spherical may acquire
the forms governed by the point-group symmetries, some being more likely
than the others -- depending on the actual number of nucleons.
In this context, it has been suggested, cf.~Ref.~\cite{DGM10}
and references therein, that each symmetry group `sufficiently rich'
in terms of symmetry elements may lead to its proper scheme of magic numbers,
which in turn can be seen as characterising such a symmetry group
from the point of view of realistic realisations
of the nuclear mean-field theories.
In particular, a series of earlier publications related to
point-group symmetries focused on the tetrahedral
and octahedral symmetry groups which are the only ones
that lead to an extra (four-fold) degeneracy of single-nucleonic levels
in {\em deformed} nuclei and to an increased nuclear stability.
In other words: The tetrahedral and octahedral symmetries are the only ones
within which both the four-fold degeneracy (`unusual' case)
appears for some levels and the two-fold degeneracy (`usual' case)
appears for some other levels. This should be compared to
the habitual two-fold (Kramers) degeneracy associated
with {\em all} the levels known in all other deformed nuclei as
e.g.~in the case of the quadrupole deformation, see e.g., Ref.~\cite{DGM10}.
In fact, the symmetry-implied large shell gaps occur
at some specific nucleon numbers and they can be comparable
to the spherical shell-gaps.  The tetrahedral magic numbers,
$N_t$ and $Z_t$, for the neutrons and protons, respectively,
are~\cite{Dudek02}: $N_t$ {\rm or} $Z_t$
= 16, 20, 32, 40, 56, 68--70, 90--94, 112,~{\rm and}~136/142.

The possibility that tetrahedral symmetry is present in atomic nuclei
has been discussed as early as in 1970'ies for $^{16}$O in relation to
the hypothetical four alpha-cluster structure~\cite{OS71,Rob78,Rob79,Rob82}.
Calculations employing the microscopic-macroscopic method~\cite{LD94},
or the Skyrme Hartree-Fock (HFB) method~\cite{TYM98,MTY98,YMM01}
suggested that, in heavier nuclei, the tetrahedral shapes
may appear in low-lying excited-, or even in the ground-states
for specific nucleon numbers;
see e.g. Refs.~\cite{OlbDoZ06,ZbMH06,ChenSG08,ZbMH09}
for more recent works.  It is, therefore, both timely and
interesting to employ the well established methods of the theory
of nuclear structure in an attempt of examining the leading features
of the excitation spectra of collective motion associated with
of the tetrahedral shape.

In this article we focus on certain properties
of nuclear rotational bands
in tetrahedral-deformed nuclei using advanced microscopic techniques
which employ the angular-momentum and particle-number projection methods.
Although the methods of performing the projections
are straightforward and well-known~\cite{RS80,HS95},
their numerical realisation is a non-trivial task,
especially for the non-axially symmetric
and non-time-reversal invariant systems
described using the cranking approximation
(see e.g.~Refs.~\cite{HHR82,ETY99}).
We have developed an efficient method to perform the projection
from general and realistic mean-field wave functions calculated
with large number of basis states~\cite{TS12}.
Characteristic features of our method which have an important impact
on an increased efficiency in the numerical realisation
of the algorithm can be summarised as follows:
(a)~An efficient truncation scheme using the information
about the occupation probabilities in the canonical basis;
(b)~The full use of the Thouless amplitude with respect to a Slater
determinant state in place of the generalised Bogoliubov amplitudes;
(c)~Avoiding the sign problem for the norm overlap
in terms of the Pfaffians~\cite{Rob09} with using the Thouless amplitude.

The tetrahedral-symmetry nuclear-states have not been so far identified
in nature.
In order to facilitate the associated experimental research program
in the case of a possible discovery of a new quantum phenomenon,
one needs to establish first of all the global and
leading characteristic properties of the tetrahedrally-symmetric nuclei.
The main focus of the present article is to test
the projection techniques associated with the mean-field methods,
as far as the characteristic features of the energy spectra
are concerned rather than trying to be
as realistic as possible in terms of the energy scale predictions.
Another aspect is to test the projection techniques necessary to calculate
the electromagnetic transition probabilities within
the nuclear-mean field theory. Such transition probabilities and/or
their branching ratios can be used as characteristic signs
of the symmetries and will become a necessary tool for
establishing such symmetries in nature. The related research program
is in progress and results will be published elsewhere.

In the next Section, Sec.~\ref{sec:method},
the principal mathematical expressions of the method are briefly recapitulated;
the details can be found in Ref.~\cite{TS12}.
The results of calculation are presented and discussed
in Sec.~\ref{sec:results}.
The final section, Sec.~\ref{sec:sum}, is devoted to the summary
and possible future perspectives.


\section{ Method of calculation }
\label{sec:method}

In what follows we assume that a nucleus is in a state corresponding to
a tetrahedral-symmetry minimum, one possibly among other competing minima
in the total energy landscape. Assuming purely static configurations,
i.e., ignoring the collective effects such as the zero-point vibrations
or any other form of, e.g., large-amplitude motion which may be particularly
needed in the case of the flat energy landscapes, we will calculate
the excitation pattern using the angular-momentum, parity,
and particle-number projection techniques,
the latter in relation to pairing, combined with the mean-field techniques.

Thus, the most general symmetry-conserving wave function
is sought in the form
\begin{equation}
      |\Psi_{M\alpha}^{I(\pm)}\rangle
      =
      \sum_{K} g_{K,\alpha}^{I(\pm)}\,
      \hat P_{MK}^I \hat P_{\pm}^{}|\Phi\rangle,
                                                             \label{eq:IpiProj}
\end{equation}
where $\hat P_{MK}^I$ and $\hat P_{\pm}$ are the angular-momentum
and parity projectors (see e.g. Ref.~\cite{RS80}).
The mean-field state $|\Phi\rangle$ is taken in the form
of the anti-symmetrised product (HFB type) wave-functions,
which is specified in Sec.~\ref{sec:meanf}.
The $K$-mixing coefficients, $g_{K,\alpha}^{I(\pm)}$, are obtained
by solving the generalised eigen-value problem of the Hill-Wheeler equations
\begin{equation}
      \sum_{K'}{\cal H}_{KK'}^{I(\pm)}\,g_{K',\alpha}^{I(\pm)}
      =
      E_\alpha^{I(\pm)}
      \sum_{K'}{\cal N}_{KK'}^{I(\pm)}\,g_{K',\alpha}^{I(\pm)},
                                                                \label{eq:HWeq}
\end{equation}
with the Hamiltonian and norm kernel matrices being defined as usual as:
\begin{equation}
      \begin{pmatrix}
         {\cal H}_{KK'}^{I(\pm)} \cr
         {\cal N}_{KK'}^{I(\pm)}
      \end{pmatrix}
      =
      \langle \Phi|\begin{pmatrix} \hat H \cr 1 \end{pmatrix}
      \hat P_{KK'}^I \hat P_{\pm}|\Phi\rangle.
                                                           \label{eq:HNkernels}
\end{equation}
In the present approach we wish to go beyond the mean-field approximation
without perturbing the tetrahedral symmetry of the problem.
This can be done by introducing in Eq.~(\ref{eq:HNkernels})
a two-body spherically-symmetric Hamiltonian $\hat{H}$,
whose form will be discussed in Sec.~\ref{sec:mhamil}.

The neutron and proton number projections require
the number projectors ($\hat P^N$ and $\hat P^Z$),
which are further included in Eqs.~(\ref{eq:IpiProj})$-$(\ref{eq:HNkernels}).
However, we found that the effect of the number projections
on the quantum spectra in the present work is small, see Sec.~\ref{sec:yb160},
and they are simply neglected in most cases
after verifying that such a neglect is justified.


\subsection{ Mean-Field Model and Its Hamiltonian }
\label{sec:meanf}

It is often of interest to employ the consistency condition between
the mean-field and the many-body Hamiltonians like,
e.g.,~in the Skyrme-Hartree-Fock models,
since it is believed that such a consistency
offers a more realistic description of the many-body systems in question.
In this article we wish to focus first of all on the nuclear excitation spectra
in the presence of pure tetrahedral symmetry of the mean-field.
In this context it is preferable to work with the model allowing
to completely control the deformation and the underlying geometrical symmetry.
For this purpose, a phenomenological deformed mean-field is more convenient.

We use the product-type states composed of the eigen-functions
of the Woods-Saxon (WS) potential~\cite{CDN87},
for which the general deformed shape is parametrised with the help
of the spherical-harmonic $\{Y_{\lambda\mu}\}$-basis:
\begin{equation}
      R(\theta,\varphi)
      =
      R_0 \,c_v(\{\alpha\})
      \bigg[
         1+\sum_{\lambda\mu}\alpha^*_{\lambda\mu}
                            Y_{\lambda\mu}^{}(\theta,\varphi)
      \bigg],
                                                              \label{eq:surf}
\end{equation}
where the coefficient $c_v(\{\alpha\})$ takes care of
the volume-conservation condition.
In the present application, the pure tetrahedral deformation
is realised by requiring all deformation parameters $\alpha_{\lambda\mu}=0$
except $\alpha_{32}$ (more precisely, $\alpha_{3+2}$ and $\alpha_{3-2}$).
In this particular case, the problem of the centre of mass
does not arise since tetrahedral-symmetric uniform-distributions preserve
the position of the centre of mass independently
of the value of the tetrahedral deformation.
The deformed WS single-particle Hamiltonian, $\hat h_{\rm def}$,
is diagonalized in the spherical harmonic oscillator basis
with the oscillator quantum numbers $n$ and $l$ satisfying
the usual relations $n_x+n_y+n_z=2n+l \le N_{\rm max}$.

The HFB type product state is obtained by further including
the monopole pairing field with the particle number constraint:
\begin{equation}
      \hat h_{\rm pair}
      =
      -
      \sum_{\tau={\rm n,\,p}} {\mit\Delta}_\tau^{}
           \left(\hat P_\tau^\dagger+\hat P_\tau \right)
      -
      \sum_{\tau={\rm n,\,p}}\lambda_\tau \hat N_\tau .
                                                            \label{eq:hpair}
\end{equation}
The pairing gap ${\mit\Delta}$ in the Hamiltonian above is
either parametrised, cf. Sec.~\ref{sec:param}, or self-consistently
calculated by using the HFB treatment assuming
the seniority interaction for neutrons and protons,
\begin{equation}
      \hat{H}_P
      =
      -\sum_{\tau={\rm n,\,p}} G_\tau^{}\, \hat P^\dagger_\tau \hat P_\tau^{}.
                                                             \label{eq:senpair}
\end{equation}
The simple-minded usage of this pairing interaction
with large model space may lead to a divergence.
We replace the monopole pairing operator, $\hat P^\dagger$,
by the one with a cut-off function, $f_{\rm c}(\epsilon)$,
\begin{equation}
      \hat P^\dagger
      =
      \sum_{i>0}d^\dagger_{i} d^\dagger_{\tilde \imath}
      \quad\rightarrow\quad
      \sum_{i>0} f_c(\epsilon_i)\, d^\dagger_{i} d^\dagger_{\tilde \imath},
                                                               \label{eq:spham}
\end{equation}
where the quantity $\epsilon_i$ and $d^\dagger_i$ are the eigen-energy
of the deformed WS single-particle state and its creation operator,
respectively, so that $\hat h_{\rm def}=\sum_i \epsilon_i d^\dagger_i d^{}_i$,
and $\tilde \imath$ refers to the time-reversal conjugate-state of $i$.
The form of the cut-off function is chosen to be~\cite{TST10},
\begin{equation}
      f_{\rm c}(\epsilon)
      =
      \frac{1}{2}
      \left[
           1+
           {\rm erf}\left(
                       \frac{ \epsilon-\lambda+\Lambda\rs{low}}{d\rs{cut}}
                    \right)
      \right]^{1/2}
      \left[
           1+
           {\rm erf}\left(
                       \frac{-\epsilon+\lambda+\Lambda\rs{up}}{d\rs{cut}}
                    \right)
      \right]^{1/2},
                                                             \label{eq:fcutoff}
\end{equation}
with the error function defined as usual by
${\displaystyle
{\rm erf}(x)=\frac{2}{\sqrt{\pi}} \int_{0}^{x}e^{-t^2}dt }$.
Following Ref.~\cite{TST10} the parameter values adopted here are:
$\Lambda\rs{up}=\Lambda\rs{low}=1.2\,\hbar \omega$ and
$d\rs{cut}=0.2\,\hbar \omega$ with $\hbar\omega=41/A^{1/3}$ MeV.
The chemical potential $\lambda$ in Eq.~(\ref{eq:fcutoff}) is
simply chosen as $(\epsilon_{i_0}+\epsilon_{i_0+1})/2$,
where $i_0$ is the last occupied orbital in the case of no pairing.
Such a treatment results in preserving typically
$+$(25-to-35) and $-$(15-to-25) states around the Fermi level
for Rare Earth nuclei depending somewhat on the deformation used.

In the present article we wish to account,
even if in a model dependent way, for at least some microscopic mechanisms
whose existence is known already. In particular we are interested in
the rotational state wave functions for increasing angular momenta.
It is known that an increase of the angular momentum causes
an increasing effect of the Coriolis coupling,
the latter resulting in a gradual increase in the rotation-induced $K$-mixing.
Since the presence of angular momentum introduces an extra direction in space,
on top of the original tetrahedral symmetry, the latter is gradually
more and more perturbed. Studying of tetrahedral symmetry of
a microscopic many-body system under the condition of increasing spin
is a matter of a compromise between the original mean-field symmetry-properties
and the Coriolis perturbation. As long as the Coriolis effects
can be considered small,
one may talk about the tetrahedral symmetry in the system.

In the pure mean-field context the rotational motion has been studied
in the past by simulating the Coriolis coupling effects
with the so-called cranking term which is linear in angular momentum
and which contains the Lagrange multiplier $\omega$ in the case of
the one-dimensional rotation. More generally,
in the case of three-dimensional rotation, as in the present case,
a triplet of Lagrange multipliers
$\{\omega_x,\omega_y,\omega_z\}\equiv\mbold{\omega}$ is introduced.
The notation can be shortened to $\mbold{\omega}=\omega \,\mbold{n}$,
where the unit vector $\mbold{n}$ points to the direction of the total spin.
One can show further that the term $\mbold\omega$ can be given
an {\em interpretation} of the rotational frequency valid asymptotically
for increasing spin and regular nuclear energy-vs.-spin dependence
-- where from the notation $\mbold\omega \to \mbold{\omega}_{\rm rot}$.

To take into account the tetrahedral symmetry mean-field, here through
$\hat{h}_{\rm def}$, the pairing correlation through the term
$\hat{h}_{\rm pair}$, which does not impact the symmetry considerations,
and a gradual effect of the Coriolis mechanism through
the $(|K|=1)$-mixing term, $\hat{h}_{|K|=1}$,
we finally introduce the mean-field Hamiltonian
\begin{equation}
      \hat h'_{\rm mf}
      =
      \hat h_{\rm def}^{}
      +
      \hat{h}_{\rm pair}^{}
      +
      \hat{h}_{|K|=1},
      \qquad
  \hat{h}_{|K|=1}\equiv-\omega_{\rm rot}\,\mbold{n}\cdot\mbold{\hat{J}}.
                                                             \label{eq:cpdefh}
\end{equation}
As it turns out the third term in the above equation with a small
$\mbold{\omega}_{\rm rot}$,
typically of the order of $\hbar\omega_{\rm rot}=0.010$ MeV,
is sufficient to break the time-reversal invariance and to introduce
the $(|{\mit\Delta K}|=1)$ $K$-mixing in the wave function,
which is important to obtain more reliably the moment of inertia~\cite{TS12}
within the Hill-Wheeler system of equations for the states
not far from the ground-state.
It has been tested, cf.~Fig.~7 in Ref.~\cite{TS12}, that the resulting spectra
do not depend very much on the particular choice of values of
$\hbar\omega_{\rm rot}$,
as long as the spin values involved are not too high.
Since in this paper we are interest in the spins of up to a dozen of $\hbar$,
the particular value of this coefficient does not play any essential role
and we keep the above value without modification for the present purpose.

Let us emphasise that the present use of the third term
in Eq.~({\ref{eq:cpdefh}}), as compared to its role
in the standard cranking model, is different.
Whereas in the cranking-model approach in numerous articles
on the high spin physics, $\omega_{\rm rot}$ plays either
the role of the Lagrange multiplier adjusted to each new spin value
and thus, on the average, increasing with spin, or, alternatively,
is currently being used as an independent cranking variable
in function of which observables such as e.g.~single-nucleon Routhians
are plotted -- here it may be seen as a coupling constant
in front of a certain phenomenological interaction term.


\subsection{ Two-Body Model Hamiltonian for Projection Calculations}
\label{sec:mhamil}

As commented already earlier, we wish to go beyond the mean-field approximation
to be able to take into account, at least partially,
certain two-body correlations which have proven to be successful
in a phenomenological description of not strongly-deformed nuclei.
For this purpose, we employ the model Hamiltonian used
in the Hill-Wheeler formalism,
cf.~Eqs.~(\ref{eq:IpiProj})$-$(\ref{eq:HNkernels}),
with an auxiliary spherically-symmetric WS potential and separable,
schematic two-body interactions as the ones employed in Ref.~\cite{TS12}.
The two-body Hamiltonian of this form does not perturb the tetrahedral symmetry,
whereas at the same time allows for including
a richer structure of the nucleon-nucleon interactions.

More specifically, we define
\begin{equation}
       \hat H=\hat h_0 + \hat H_F + \hat H_G,
                                                                \label{eq:hamT}
\end{equation}
where $\hat h_0$ is a one-body Hamiltonian composed of
the kinetic energy term and the spherical WS potential
(with the Coulomb interaction for protons).
The schematic particle-hole ($F$-type) interaction,
$\hat H_F$, is chosen to be isoscalar and is defined by
\begin{equation}
      \hat H_F
      =
      -\frac{1}{2}\,\chi\sum_{\lambda\ge 2}
       \sum_\mu :\hat F^\dagger_{\lambda\mu} \hat F_{\lambda\mu}:,
       \qquad
       \hat F_{\lambda\mu}=\sum_{\tau={\rm n,\,p}} \hat F_{\lambda\mu}^\tau,
                                                                \label{eq:hamF}
\end{equation}
where $:\ :$ denotes the normal ordering and $\tau={\rm n,\,p}$
distinguishes neutrons and protons.
Furthermore, the spatial representation of
the above particle-hole type operator, $\hat F_{\lambda\mu}^\tau$,
is defined through the one-body field,
\begin{equation}
       F^\tau_{\lambda\mu}(\bm{r})
       =
       R^\tau_0\, \frac{dV_c^\tau}{dr}\,
                  Y_{\lambda\mu}(\theta,\phi),
                                                              \label{eq:hamFop}
\end{equation}
with $V_c^\tau(r)$ and $R^\tau_0$ being the central part of the WS potential
and its radius, respectively.
The so-called self-consistent value~\cite{BM75} of the force strength,
$\chi$, common to all multipolarities, is calculated by
\begin{equation}
      \chi=(\kappa_n+\kappa_p)^{-1},
      \qquad
      \kappa_\tau\equiv \left(R_0^\tau\right)^2
      \int_0^\infty
      \rho_0^\tau(r)\frac{d}{dr}\left(r^2\frac{dV^\tau_c(r)}{dr}\right) dr,
                                                                \label{eq:strF}
\end{equation}
where $\rho_0^\tau(r)$ is the density of a hypothetical spherical ground state,
which is calculated with the filling approximation for
each nucleus based on the spherical WS single-particle state of $\hat h_0$.
On the other hand, the pairing type ($G$-type) interaction,
$\hat H_G$, acts only within like-particles, and is given by
\begin{equation}
      \hat H_G
      =
      -\sum_{\tau={\rm n,\,p}}\sum_{\lambda\ge 0} g^\tau_\lambda\,
       \sum_\mu\hat G^{\tau\dagger}_{\lambda\mu} \hat G^\tau_{\lambda\mu},
       \qquad
      \hat G^{\tau\dagger}_{\lambda\mu}\equiv\frac{1}{2}\sum_{ij}
      \langle i| G^{\tau}_{\lambda\mu} |j \rangle
       c^\dagger_i c^\dagger_{\tilde j},
                                                                \label{eq:hamG}
\end{equation}
where the matrix elements of the pairing type operator,
$\hat G^{\tau\dagger}_{\lambda\mu}$,
are calculated with the help of the standard multipole form,
\begin{equation}
       G_{\lambda\mu}(\bm{r})=\left(\frac{r}{\bar R_0}\right)^\lambda
       \sqrt{\frac{4\pi}{2\lambda+1}}\, Y_{\lambda\mu}(\theta,\phi),
                                                                 \label{eq:opG}
\end{equation}
with $\bar R_0=1.2 A^{1/3}$ fm.

The present formalism follows the main lines of Ref.~\cite{TS12}
with a few modifications.
Firstly, the extra one-body terms ($\hat h_1=-\hat h_F - \hat h_G$
in \S3.1 in \cite{TS12}) are included in Ref.~\cite{TS12}
in order to cancel out the one-body exchange contributions
of the multipole interactions, $H_F+H_G$.
It turns out that the effect of these terms on
the resultant projected spectra is small,
so that they are neglected for simplicity.

A slightly different deformed mean-field Hamiltonian
has been used in Ref.~\cite{TS12},
namely the one derived as the Hartree approximation
to the interaction~(\ref{eq:hamF}), in the form
$\hat h_{\rm def}=\hat h_0
- \sum_{\lambda\mu}\alpha_{\lambda\mu}{\hat F}_{\lambda\mu}$,
which, however, coincides with
the central part of the present deformed WS potential only within
the first order in the deformation parameters $\{\alpha_{\lambda\mu}\}$.
We employ, in the present work, the general shape parameterisation
based on the deformed radius in Eq.~(\ref{eq:surf})
with the volume-conservation condition properly taken into account.
Moreover, the cut-off of the pairing model space is introduced
directly in the operator $\hat G^{\tau\dagger}_{\lambda\mu}$ in
Ref.~\cite{TS12}; then one has to use the spherical single-particle
energy in the cut-off function~(\ref{eq:fcutoff})
to keep the spherical invariance of the Hamiltonian.
We found that it sometimes causes a problem that the results
are rather sensitive to the choice of model space
for the relatively small pairing model space like in the present case,
$\Lambda\rs{u}=\Lambda\rs{l}=1.2\,\hbar \omega$.
In the present calculation, the cut-off of the pairing model space
is taken into account in the step of
deformed HFB calculation in Eqs.~(\ref{eq:cpdefh})$-$(\ref{eq:spham})
based on the deformed WS single-particle state.
Therefore, the cut-off function is not included explicitly
in the pairing operator $\hat G^{\tau\dagger}_{\lambda\mu}$ anymore.


\subsection{ Choice of Parameters}
\label{sec:param}

The deformed mean-field single-particle states are calculated
in the present work using the Woods-Saxon potential.
An often used parameterisation introduced over thirty years back
is referred to as `universal' (cf. Refs.~\cite{DSW81,DW78,DMS79,KRD89,LPD95}).
We employ, in this work, a new improved ``universal compact'' set,
whose parameters are listed in Table~\ref{tab:WSunivcompact},
see Ref.~\cite{CDN87} for notations.

\begin{table*}[hbtp]
\caption{ The parameters of the Woods-Saxon potential used in this work.
Symbols $\nu$ and $\pi$ refer to neutrons and protons, respectively.
}
\label{tab:WSunivcompact}
\begin{ruledtabular}
\begin{tabular}{c|ccccccccc}
&$V_{\rm 0c}$ [MeV] & $\kappa_{\rm c}$ & $r_{\rm 0c}$ [fm] & $a_{\rm c}$ [fm] &
$\lambda$ & $V_{\rm 0so}$ [MeV] & $\kappa_{\rm so}$ & $r_{\rm 0so}$ [fm] &
$a_{\rm so}$ [fm] \\
 \hline
\ $\nu$  & $-$52.0 & 0.650 & 1.26 & 0.64 & 28.0 & 49.6 & 0 & 0.870 & 0.70 \\
\ $\pi$  & $-$53.0 & 0.526 & 1.27 & 0.71 & 23.0 & 49.6 & 0 & 0.888 & 0.86 \\
\end{tabular}
\end{ruledtabular}
\end{table*}

As for the maximum number of the harmonic oscillator shells to be used
in the calculation, we employ $N_{\rm max}=20$,
which is a safe margin to accurately calculate the single-particle
wave functions of the bound-states in the Woods-Saxon potential,
and at the same time guarantees the convergence
of the result of projection calculations~\cite{TS12}.

The ground state deformation is determined
for each nucleus by the axially symmetric WS-Strutinsky calculation
of Ref.~\cite{TST10}, where an algorithm allowing for the treatment
of nuclei with weakly bound nucleons has been implemented.
In the present realisation we calculate the strength of the seniority force,
$G$, so as to reproduce the even-odd mass differences
with the calculated deformation.
More precisely, for the calculated equilibrium (ground state) deformations,
$\alpha_{20}$ and $\alpha_{40}$, we adjust the $G$-values
in such a way that the calculated pairing gap,
$\Delta_\tau=G_\tau\,\langle \hat P^\dagger_\tau \rangle$ ($\tau$=n,\,p),
agrees with the even-odd mass difference.
Once the parameters $G$ are fixed in this way,
the usual BCS or the HFB equations (in the case of the cranking Hamiltonian)
are solved self-consistently at any given deformation.

As for the spherically-symmetric Hamiltonian used
for the projection calculation, the self-consistent value $\chi$
for the $F$-type interaction in Eq.~(\ref{eq:strF})
is used without any modifications.
We include the $\lambda=2,3,4$ components for the $F$-type
interaction in Eq.~(\ref{eq:hamF}),
because the $\alpha_{20}$ and $\alpha_{40}$ deformations are taken into account
for the ground state within the Strutinsky method
and the tetrahedral shape is described by the $\alpha_{32}$ deformation.
As for the $G$-type interaction, the monopole pairing is known to be essential.
It has been recognised that the quadrupole pairing interaction
is also important especially to describe the rotational motion~\cite{BM75}.
Therefore we include the $\lambda=0,2$ components for
the $G$-type interactions in Eq.~(\ref{eq:hamG}).
The strength of the monopole pairing is determined again
to reproduce the pairing gap, i.e.,
$\Delta_\tau=g_0^\tau\,\langle \hat G^{\tau\dagger}_{00} \rangle$
($\tau$=n,\,p), for the ground state wave function.
As for the strength of the quadrupole pairing
we assume $g^\tau_2/g^\tau_0=13.6$,
which is determined to approximately reproduce the moment of inertia of
the ground state bands in the previous calculation~\cite{TS12}.
Note that the values of $G_\tau$ and $g^\tau_0$ are slightly different,
because the monopole pairing operator $\hat P^\dagger_\tau$
and the associated strength, $G_\tau$, are defined with respect to
the deformed WS basis, while the corresponding operator
$\hat G^{\tau\dagger}_{00}$ and its strength,
$g^\tau_0$, are defined with respect to the spherical WS basis.

The model space truncation in the projection calculation is controlled
by the small parameter $\epsilon$ defined through the requirement, that
the orbitals which in canonical representation
have occupation probabilities $v_i^2>\epsilon$ are included,
and, similarly, the core orbitals with $1-v_i^2 < \epsilon$.
We have chosen $\epsilon=10^{-6}$ in the present calculation,
with which it is confirmed that the resultant projected energies
are stable within in 1 keV, which corresponds to six digits of accuracy
for absolute energy of the present Hamiltonian in Eq.~(\ref{eq:hamT}).
As for the calculation of the Hill-Wheeler Eq.~(\ref{eq:IpiProj}),
the states that have smaller norm eigen-values than $10^{-10}$
are excluded.


\section{ Results of the Calculations }
\label{sec:results}

According to the calculations in Refs.~\cite{LD94,Dudek02},
the tetrahedral magic numbers are $N_t$ or $Z_t$
= 16, 20, 32, 40, 56\,--\,68, 70, 90\,--\,94, 112, and 136/142,
and are the same for the neutrons and protons.
Calculations which followed, Refs.~\cite{Dudek09}, figure~3,
and \cite{Schun04}, using the universal Woods-Saxon mean-field Hamiltonian,
suggested that, in particular, $^{80}_{40}$Zr$_{40}$ and $^{96}_{40}$Zr$_{56}$
are tetrahedrally-symmetric in their ground-states,
whereas tetrahedral minima lie about 1 MeV above the ground-state
in $^{110}_{\;\;40}$Zr$_{70}$.  Skyrme Hartree-Fock calculations
for the latter nucleus predict the possibility of the tetrahedral minima
being the lowest
-- depending strongly on the choice of the model parameters~\cite{OlbDoZ06}.
More recently, we have reported on the calculations of
the tetrahedral spectra in $^{108,110}$Zr, in Ref.~\cite{YKIS2011}.
The predicted shape coexistence in the Zirconium region
which includes tetrahedral symmetry minima may give rise to,
among others, the presence of isomeric states.
Quite recently the experiment have been performed for unstable
nucleus $^{108}$Zr and the results are compatible
with the existence of an isomeric state~\cite{Sumik11}
whose nature is being debated.

\subsection{Remarks about Symmetry Properties and Quantum Rotors}

One of our goals is to determine whether the microscopic calculations
which combine various advanced techniques of the nuclear quantum mechanics
reproduce the excitation pattern predicted by group-representation theory.
Because all of the two-fold and four-fold degenerate single-particle states
are occupied with equal probabilities for the doubly-closed shell
tetrahedral configurations, the totally symmetric,
so-called {\em tensor} $A_1$ irreducible representation
of the tetrahedral point-group $T_d$
(as opposed to the {\em spinor} irreducible representations characterising
the symmetry properties of the single-nucleonic wave-functions
within the tetrahedral double-point group $T_d^D$,
cf.~also Table VIII in Appendix) can be expected
as the resulting symmetry of the full system in its lowest rotational band.
The situation remains the same if the seniority-type pairing interaction
is effective for fully paired even-even nuclei.

An ideal tetrahedral ($T_d$\,-symmetric) classical rotor is often referred
to as `spherical', because its moment of inertia tensor is diagonal
with all elements strictly equal. In the case of a quantum rotor,
the notion of the inertia tensor cannot be strictly-speaking defined
since the only quantum observables directly associated with rotational motion
of such an object are the energy and angular momentum.
The specific spectral properties of {\em quantum rotors with
point-group symmetries} have been actively studied in relation
to the TetraNuc Collaboration activities in recent years.
For instance, examples of the octupole-symmetric quantum-rotor spectra
have been presented in Ref.~\cite{JDu01};
similar examples for specifically tetrahedral-symmetric quantum rotors
can be found in Ref.~\cite{JDu07} whereas the underlying tensor formalism
and the general form of the reduced matrix elements are presented
in Ref.~\cite{AGo08}. 
Observe that in contrast to the `usual' quantum rotor Hamiltonians
discussed in the literature, which are quadratic forms of
the angular momentum operators, the tetrahedral
(or other octupole-symmetric) rotor Hamiltonians are
specific third-order forms expressed in terms of the operators 
$\{\hat{I}_x,\hat{I}_y, \hat{I}_z\}$ (equivalently of 
$\{\hat{I}_{-1},\hat{I}_0,\hat{I}_{+1}\}$
using spherical tensor representation). 
Furthermore, relations between the energy spectra of the quantum rotors
and the associated properties of classical rotors have been
discussed in Ref.~\cite{MMi04}.

It can be shown that the spectra of tetrahedral-symmetric structure-less
quantum rotor are composed of the $(2I+1)$-degenerate states
for each given spin $I$, cf.~Ref.~\cite{JDu07}.
In this sense the spherical-symmetry of the classical tetrahedral rotor
mentioned above and the symmetry of the tetrahedral quantum rotor
can be seen as analogous. According to group-theory considerations,
each of the $(2I+1)$-degenerate states of the $T_d$\,-symmetric
structure-less rotor of any given $I$ belongs to a certain specific
irreducible representation of the group in question.
Among five of those irreducible representations,
the $A_1$ (scalar) representation contains states
with the following characteristic set of spin-parity combinations
[cf. Table VI in the Appendix (also e.g.~Ref.~\cite{HerzbII,HerzbIII})]:
\begin{equation}
 0^+,\,3^-,\,4^+,\,6^+,\,6^-,\,7^-,\,8^+,\,9^+,\,9^-,\,10^+,\,10^-,
 \,11^-,\,2 \times 12^+,\,12^-,\cdots.
\label{eq:TdA1}
\end{equation}
As it is discussed in more detail below,
such characteristic spectra are indeed realised for the lowest energy band
in the results of our microscopic calculation, cf.~also Ref.~\cite{YKIS2011}.

It is worthwhile mentioning that there exist certain extra discrete symmetries
for the tetrahedral shape nuclear mean-field configurations applying within
the cranking model. They are referred to as {\em doublex} and {\em triplex}
(as opposed to the `usual' {\em simplex-symmetry} applying to
the cranking model for the pear-shape symmetric nuclei).
The corresponding quantum numbers are useful to further classify
the characteristic spectra as discussed in Refs.~\cite{Frau01,NSc05},
but this issue goes beyond the scope of the present article.

In this work we report on the more extensive and detailed investigations
for three doubly-closed tetrahedral-shell configurations in nuclei:
$^{160}$Yb ($Z=70$ and $N=90$), the already mentioned $^{110}$Zr,
as well as the heavier tetrahedral-symmetric nucleus
in the Actinide region, $^{226}$Th ($Z=90$ and $N=136$).
The axially symmetric octupole ($\alpha_{30}$) deformed states
of the latter nucleus have also been studied  in Ref.~\cite{TS12}.


\subsection{ Tetrahedral-Symmetry in $^{160}$Yb:
 Phonon-vs.-Rotation-Like Structures}
\label{sec:yb160}

Following the procedure of Sec.~\ref{sec:param},
we obtain the following parameters for the pairing force
and residual interaction given in Table~\ref{tab:Yb},
where the calculated deformation parameters for the ground state
are also shown.

\begin{table}[ht]
\begin{center}
\begin{tabular}{ccccccccc}
\hline
$\alpha_{20}$ & $\alpha_{40}$ &
${\mit\Delta}_{\rm n}$ [MeV] & ${\mit\Delta}_{\rm p}$ [MeV] &
$G_{\rm n}$ [MeV] & $G_{\rm p}$ [MeV] &
$\chi$ [MeV$^{-1}$] & $g_0^{\rm n}$ [MeV] & $g_0^{\rm p}$ [MeV] \cr
\hline
0.194\  & 0.031\  & 1.265 & 1.370 &
0.1512 & 0.1732 & 2.643$\times10^{-4}$ & 0.1468 & 0.1675 \cr
\hline
\end{tabular}
\end{center}
\caption{
The calculated ground state deformation parameters
($\alpha_{20}$, $\alpha_{40}$),
the 4-th order even-odd mass difference
$({\mit\Delta}_{\rm n}$, ${\mit\Delta}_{\rm p}$),
and the force strength parameters determined based on them
for $^{160}_{\;\;70}$Yb$^{}_{90}$.
The value $g_2^\tau/g_0^\tau=13.6$ is taken for
the ratio of the quadrupole and monopole pairing.
}
\label{tab:Yb}
\end{table}

\begin{figure}[!ht]
\begin{center}
\includegraphics[width=70mm]{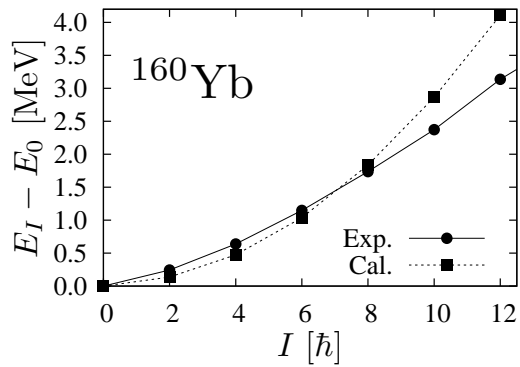}
\vspace*{-4mm}
\caption{
Comparison of the experimental and the calculated
rotational spectra in $^{160}$Yb.
}
\label{fig:YbEI}
\end{center}
\end{figure}

Let us begin by presenting the results of our projection calculations
for the ground state band in $^{160}$Yb; the corresponding results are shown
in Fig.~\ref{fig:YbEI}, where comparison with the experimental data
can also be found.
In this calculation, we use the HFB type wave function
with a small Coriolis $(|\Delta K|=1)$-coupling
$\hbar\omega_{\rm rot}=0.010$ MeV parameter,
as discussed in Sec.~\ref{sec:meanf}.
We choose the cranking axis as the $y$-axis,
which is perpendicular to the symmetry axis ($z$-axis).
The agreement of the calculated spectra with the experimental data
is acceptable but discrepancies as compared to the experiment
increase with spin.
In particular, the observed moment of inertia increases as a function of spin,
but the calculated moment of inertia is fairly constant.
This trend was already found for other nuclei in Ref.~\cite{TS12}
and may partly reflect the fact that in contrast to the cranking model,
within which the pairing correlations systematically decrease
with cranking frequency and thus increase the moment of inertia,
in the present model with projection from one mean-field state
the mentioned mechanism does not exist.
In the present calculation of the ground state the axial symmetry
is broken only by a small $(|\Delta K|=1)$-mixing term,
and the mixing effect of the $K$ quantum number
in the Hill-Wheeler Eq.~(\ref{eq:HWeq}) remains small.
Since this particular aspect is of secondary importance of the present project
we accepted the disagreement in question as remaining under control
but without consequences for the main conclusions.
Doing so we may profit from a technical advantage:
It is sufficient to use relatively small numbers of points
for Gauss quadratures with respect to the Euler angles
when performing the angular-momentum projection calculation,
e.g., $N_\alpha=N_\gamma=16$ and $N_\beta=50$.

We proceed to examining the tetrahedral nuclear configuration
and related rotational states.
The simplest way to construct the tetrahedral shape
in the surface parameterisation in Eq.~(\ref{eq:surf})
is to set $\alpha_{32}=\alpha^*_{32}=\alpha_{3-2}$
as the only non-zero deformation parameters.
In the coordinate system chosen
the upper ($z>0$) and lower ($z<0$) sides of the tetrahedron
are parallel to the $x$- and $y$- axes, respectively, and
the $z$-axis is along the line that connects
the middle points of these two facing sides.
Again, we have chosen the $y$-axis for cranking with a small frequency
$\hbar \omega_{\rm rot}=0.01$ MeV.
Obviously, in this case the axial symmetry is strongly broken
depending on $\alpha_{32}$ and thus
higher order quadratures in the projection calculations are necessary.
We take $N_\alpha=N_\gamma=N_\beta=64$ for $\alpha_{32}=0.1-0.2$,
$N_\alpha=N_\gamma=N_\beta=84$ for $\alpha_{32}=0.25-0.30$,
and $N_\alpha=N_\gamma=N_\beta=104$ for $\alpha_{32}=0.35-0.40$
for the calculation in this nucleus.

\begin{figure}[!ht]
\begin{center}
\includegraphics[width=160mm]{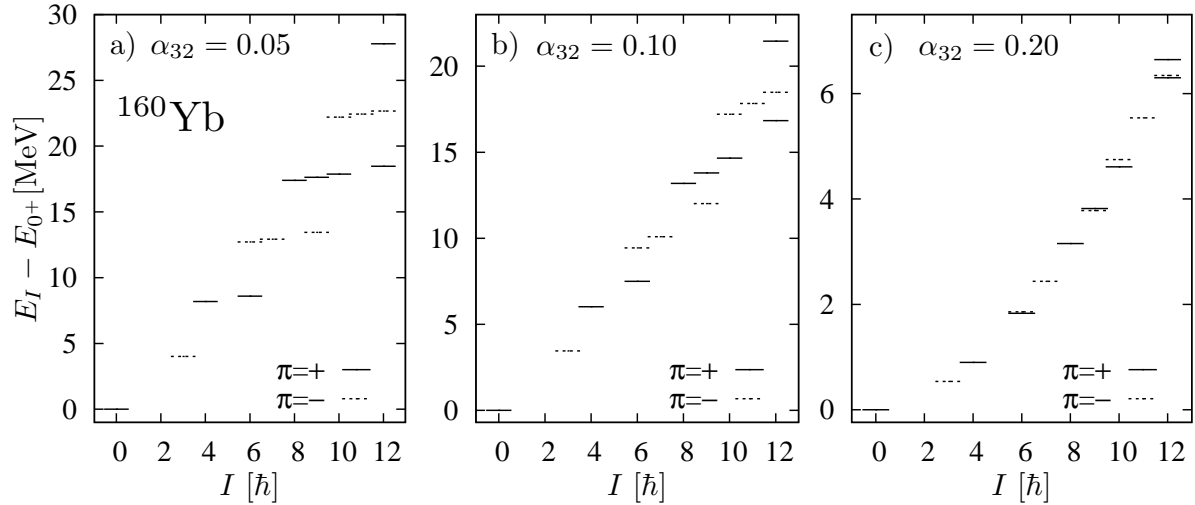}
\vspace*{-4mm}
\caption{
Examples of calculated spectra of tetrahedral states
belonging to the lowest energy part in the spectrum in $^{160}$Yb with
$\alpha_{32}=0.05$~(left), $\alpha_{32}=0.10$~(middle)
and $\alpha_{32}=0.20$~(right),
testing the dependence of the excitation energies
on the tetrahedral deformation [cf.\,also Eq.~(\ref{eq:TdA1})].
The positive (negative) parity states are denoted by the solid (dotted) lines.
Note the difference of ordinate-scales between the left and right panels.
}
\label{fig:Ybt10}
\end{center}
\end{figure}

In Figure~\ref{fig:Ybt10} we illustrate the result of excitation energies
of the projected eigen-states, cf. Eq.~(\ref{eq:IpiProj}), whose norm
is not very small, i.e. not less than $10^{-5}$ of that of the ground state.
They compose the lowest-energy sequence,
and calculated for the pure tetrahedrally deformed configuration
with deformations $\alpha_{32}=0.05$, $0.10$ and~$0.20$.
As it is seen from the figure, only the specific spin-parity combinations
appear, cf. Eq.~(\ref{eq:TdA1}), for all deformations.
This pattern is characteristic for the tetrahedral symmetry
and is quite different from the one of the usual quadrupole deformation.
Indeed, the spectrum is composed of states characterised by
$0^+$, $3^-$, $4^+$, $6^{\pm}$, ...;
the states with $I=1,\,2$ and~5 are missing in the figure
since their norms are too small and/or they lie much higher
in energy as compared to the discussed lowest-energy sequence.

A very small tetrahedral deformation of 
$\alpha_{32}=0.05$ corresponds to nearly spherical form.
The existence of collective excitations in spherically symmetric nuclei
has long been recognised in terms of the vibrational modes,
which results in an equidistant spectrum composed of multiplet of states,
within the simplest harmonic representation,
in terms of the vibration-quanta:
The phonons, see e.g.~Sect.\,(6.3.2) in Ref.~\cite{WGr96}.

The spectrum with smallest deformation $\alpha_{32}=0.05$
in Fig.~\ref{fig:Ybt10} is more vibrational-like,
while that with $\alpha_{32}=0.20$ is
approaching to the rotational-like spectrum;
the one with $\alpha_{32}=0.10$ is in-between.
In fact, ($4^+,6^+$), ($6^-,7^-,9,^-$), ($8^+,9^+,10^+,12^+$)~... states
in Fig.~\ref{fig:Ybt10}~a) having the same parity can be grouped together,
and would be interpreted as slightly perturbed
two-phonon, three-phonon, four-phonon~... multiplet structures, respectively,
of an elementary mode of the $3^-$ vibrational excitation.
Moreover, the excitation energies of $3^-$, $6^+$, $9^-$, $12_1^+$ form
a rather linear dependence as a function of spin;
the dependence resembles the pattern expected for the multi-phonon excitation.
Note, however, that only the specific spin states
among the multi-phonon multiplets appear,
which is a consequence of the tetrahedral symmetry.
In contrast in Fig.~\ref{fig:Ybt10}~c),
the states with the same spin value are nearly degenerate,
which is a specific feature of the ideal rotor,
and, at the same time, approximately follow
the quadratic energy-vs.-spin relation, $E(I) \propto I(I+1)$.

Let us emphasise, that both the parity and the angular-momentum
projections were essential for obtaining the tetrahedral-symmetry pattern
predicted by the group theory.
This symmetry pattern seems rather typical for the present model Hamiltonian:
We obtain similar pattern also for other nuclei,
e.g., $^{110,108}$Zr~\cite{YKIS2011} and $^{226}$Th below.
In the present work, we concentrate on the lowest energy sequence and
we do not enter the discussion of the group theory aspects.
Instead let us only mention that representations other than
$A_1$ must be expected for excited bands;
also -- one may expect, that the symmetry in the case of
the non-doubly-closed shell nuclei could be manifested less strongly.

\begin{figure}[!htb]
\begin{center}
\includegraphics[width=130mm]{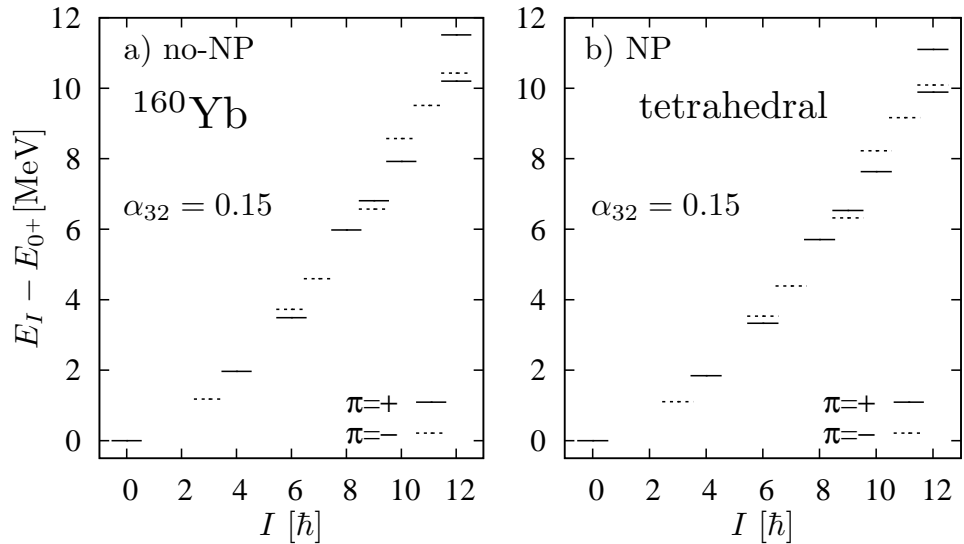}
\vspace*{-4mm}
\caption{
Comparison of the spectra of tetrahedral states
with (right) and without (left) the number projection (NP),
which are shown in the same way as in Fig.~\ref{fig:Ybt10}
but with $\alpha_{32}=0.15$.
}
\label{fig:Ybt10NP}
\end{center}
\end{figure}

In Figure \ref{fig:Ybt10NP} we compare the results of tetrahedral spectra
with and without the particle-number projection (NP) related
to pairing formalism.
Although the moments of inertia (the slopes) are slightly different,
the characteristic properties of the spectra are exactly the same
in the two calculations.
We conclude that the effect of the particle number projection is small
and we do not apply it in the rest of the article.


\subsection{Transition to Ideal Rotor and Moments of Inertia in $^{160}$Yb}
\label{sec:momiyb}

\begin{figure}[!ht]
\begin{center}
\includegraphics[width=140mm]{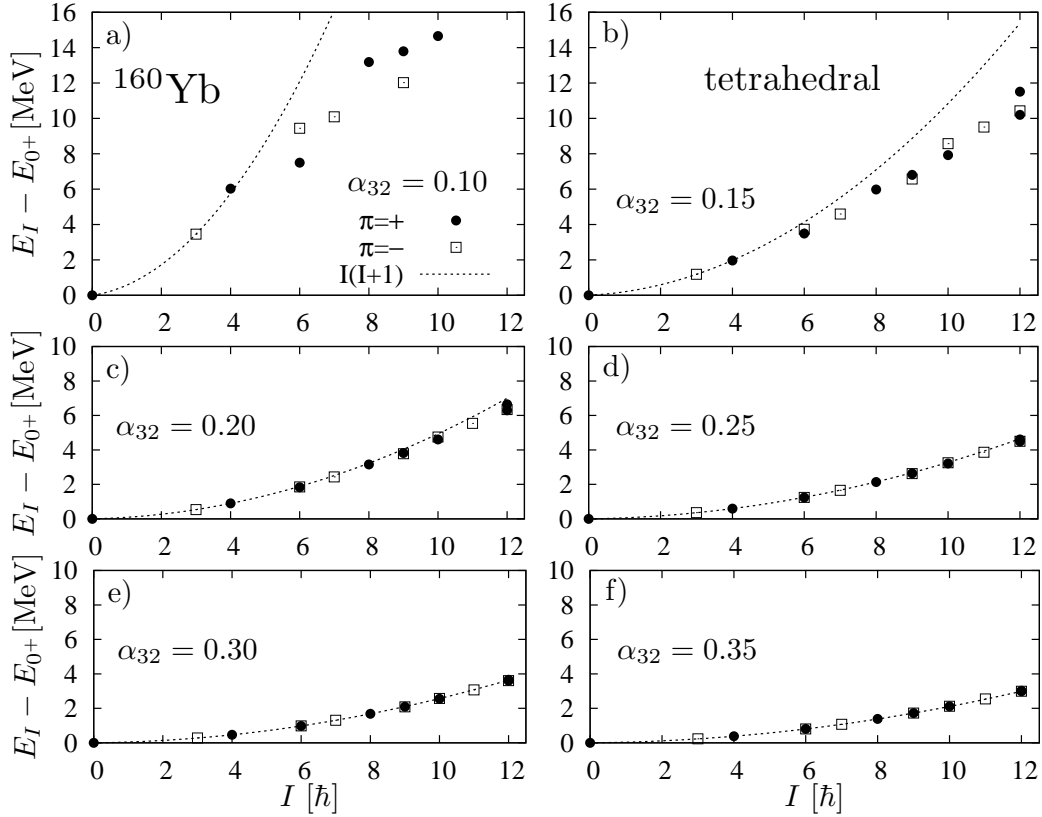}
\vspace*{-4mm}
\caption{
Calculated spectra of tetrahedral states in $^{160}$Yb
with $\alpha_{32}=0.10$, 0.15, 0.20, 0.25, 0.30, and 0.35,
respectively, for a), b), c), d), e), and f).
The dotted line in each panel denotes an ideal $I(I+1)$ sequence
going through the first excited $3^-$ state.
Note that almost exact degeneracies for $I=(6^+,6^-),
(9^+,9^-),(10^+,10^-),(2 \times 12^+,12^-)$ states
are obtained for $\alpha_{32}\ge 0.25$ demonstrating
the nearly perfect rotor character of the rotational excitation of the system.
}
\label{fig:Ybtall}
\end{center}
\end{figure}

The fact that the tensor of inertia of an ideal classical tetrahedral rotor
is diagonal with all components equal (`spherical rotor')
suggests that the tilting direction of the cranking axis
in Eq.~(\ref{eq:cpdefh}) may not affect the tetrahedral spectra,
at least to the extent in which the Coriolis alignment effects
can be neglected, i.e., for not too high spins.
In order to test this conjuncture, we investigated the projected spectra
from the cranked mean-field state with the different tilted cranking axes.
We have varied the cranking axis in our coordinate system, i.e.
the vector $\mbold{n}$ in the term,
$\hat{h}_{|K|=1}\equiv-\omega_{\rm rot}\,\mbold{n}\cdot\mbold{\hat{J}}$,
in Eq.~(\ref{eq:cpdefh}) is changed by
\begin{equation}
    \mbold{n}=(\sin\theta\sin\varphi,\cos\theta,\sin\theta\cos\varphi),\qquad
	0 \le \theta \le 90^\circ,\quad 0 \le \varphi \le 45^\circ.
                                                           \label{eq:tiltangle}
\end{equation}
We found that the differences of the resulting spectra
for the lowest energy sequence are negligible within the accuracy
of our calculation;
i.e. the nature of `spherical rotor' is numerically confirmed.
More generally, the projected spectra for the tetrahedral symmetric nuclei
do not depend on the $|\Delta K|=1$ coupling term,
both the strength $\omega_{\rm rot}$ and
and the direction $\mbold{n}$ of the tilted cranking axis,
as long as $\omega_{\rm rot}$ is small.

\begin{figure}[!ht]
\begin{center}
\includegraphics[width=140mm]{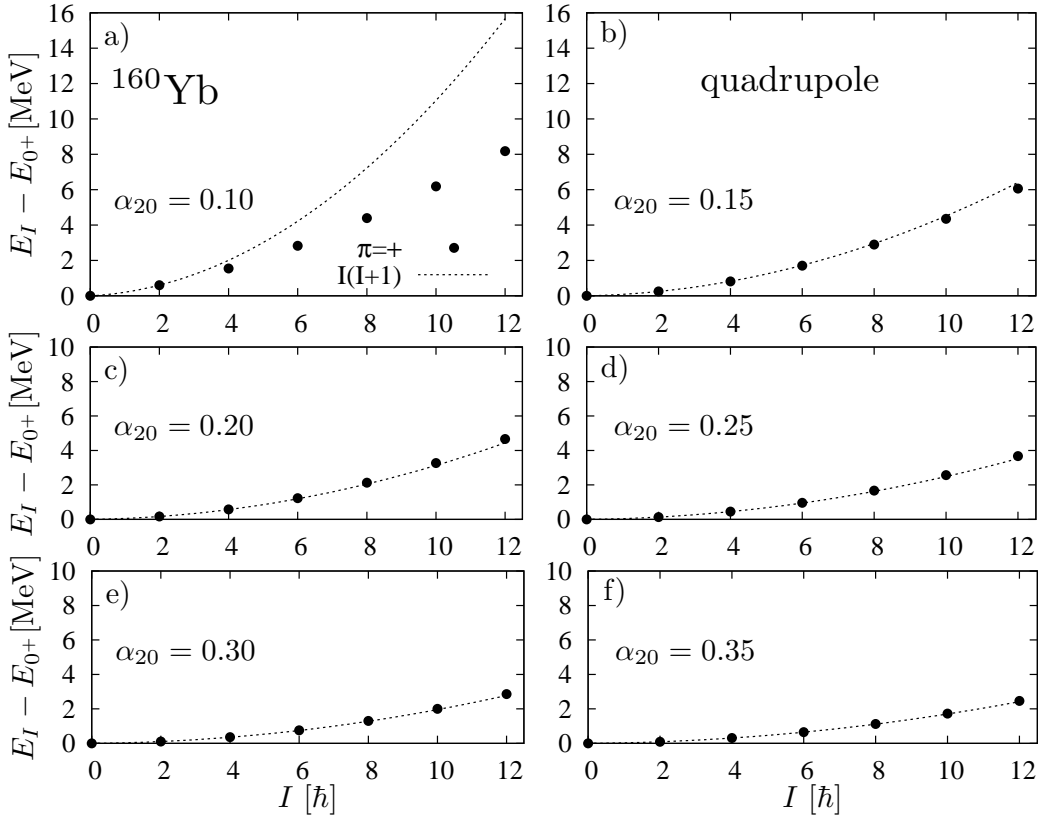}
\vspace*{-4mm}
\caption{
Calculated spectra of quadrupole deformed states in $^{160}$Yb
with $\alpha_{20}=0.10$, 0.15, 0.20, 0.25, 0.30, and 0.35, respectively,
for a), b), c), d), e), and f).
The dotted lines in each panel denote the ideal $I(I+1)$ spectra
going through the first excited $2^+$ state.
}
\label{fig:Ybqall}
\end{center}
\end{figure}

In Figure~\ref{fig:Ybtall} we show the calculated excitation energies
for selected values of the tetrahedral deformation.
The ideal rotor spectra with the energies proportional to $I(I+1)$
and containing the calculated first excited $3^-$ states
are also shown by the dotted lines.
This figure clearly shows that the spectra exhibit the gradual transition
from linear to parabolic spin dependence
with increasing the tetrahedral deformation:
The almost ideal rotor spectrum is realised for $\alpha_{32} \gtsim 0.25$.

Now we compare the tetrahedrally-symmetric spectra with those
of the quadrupole deformation.
In Figure~\ref{fig:Ybqall} we show the results of calculated spectra
obtained by the angular-momentum projection from the pure quadruple
deformed states, where all the deformation parameters are set to zero
except $\alpha_{20}$ (no parity projection is required in this case).
The projection calculation tends to give good rotational spectra,
but the result with small deformation, $\alpha_{20}=0.10$,
considerably deviates from the one for the pure rotor spectra.
Thus the gradual transition from the linear to parabolic energy-vs.-spin
dependence is seen also for the calculation of the quadrupole deformation.

\begin{figure}[!ht]
\begin{center}
\includegraphics[width=140mm]{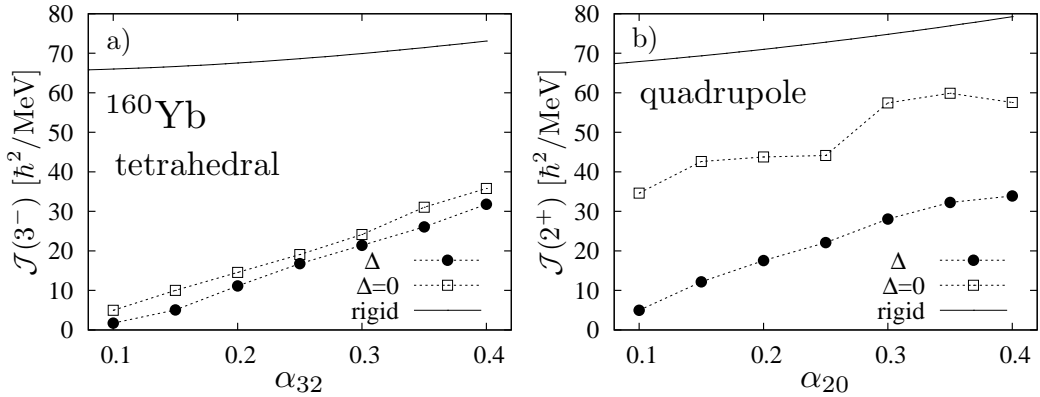}
\vspace*{-4mm}
\caption{
Moment of inertia estimated from the calculated spectra
for the pure tetrahedral states (left) and the pure quadrupole states (right)
as functions of the deformation parameters in $^{160}$Yb.
The energy of the first excited $3^-$ ($2^+$)
is used for estimation of the former (latter).
The results with the pairing correlations artificially set to zero
are also included.
The classical rigid-body moments of inertia in function of
the deformation parameters are shown as solid lines.
}
\label{fig:Ybmoi}
\end{center}
\end{figure}

\begin{figure}[!ht]
\begin{center}
\includegraphics[width=140mm]{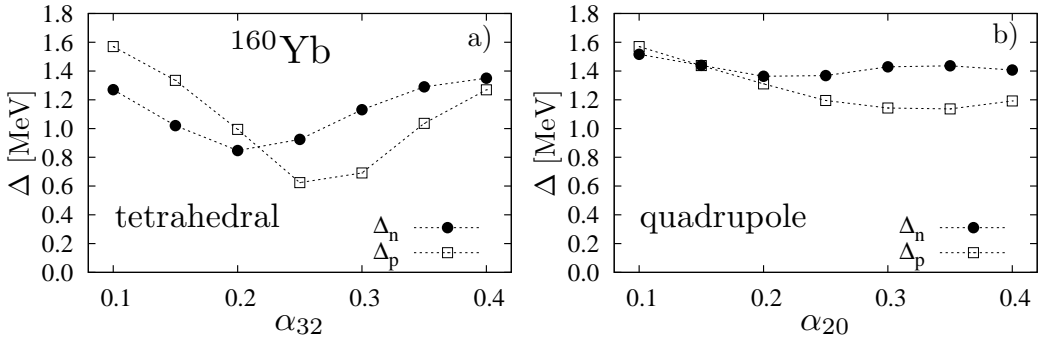}
\vspace*{-4mm}
\caption{
The self-consistent neutron and proton pairing gaps
for the tetrahedral states (left) and the quadrupole states (right)
as functions of the respective deformation parameters in $^{160}$Yb.
}
\label{fig:Ybdel}
\end{center}
\end{figure}

Although the moment of inertia is {\em not} any quantum-mechanical observable,
certain quasi-classical analogies often found in the literature allow
to define and estimate the corresponding values.
Here we define this parameter, ${\cal J}$, through
\begin{equation}
       E(I)-E(0)=\frac{I(I+1)}{2{\cal J}}.
                                                              \label{eq:rotEI}
\end{equation}
It is well-known that the moments of inertia of observed rotational band
near the ground state are about (or even smaller than) half of
the classical rigid-body value.  This large reduction is supposed
to be due to the pairing correlations~\cite{BM75,MdV83}.
In fact, the moments of inertia extracted from the high-spin states,
where the pairing correlations are believed to be quenched,
are known to be close to the rigid-body value,
although some deviations attributed shell effects exists,
see e.g.~Ref.~\cite{DFP04}.
Therefore, it is instructive to investigate the moment of inertia
in the case of the tetrahedral rotor.
In Figure~\ref{fig:Ybmoi} the moments of inertia calculated
from Eq.~(\ref{eq:rotEI}) are plotted as functions of the tetrahedral (left)
and the quadrupole (right) deformations, where they are estimated
from the calculated $3^-$ and $2^+$ excitation energies, respectively.
The results with neglecting the pairing correlation
and the rigid-body value are also included.
An irregular behaviour for the unpaired ($\Delta=0$) moments of inertia,
i.e., at $\alpha_{20}\approx 0.20-0.25$,
is due to the fact that the level crossings near the Fermi surface occur.
In order to illustrate the possible correlation between the moments of inertia
and the intensity of pairing correlations measured
with the help of the pairing gaps, the calculated pairing gaps
at corresponding deformations are shown in Fig.~\ref{fig:Ybdel}.

As it becomes clear from Fig.~\ref{fig:Ybmoi},
the moment of inertia increases rapidly with increasing deformation
for both the tetrahedral and quadrupole shapes, indicating that
the picture of the good rotor emerges for larger deformation.
Observe that the values of the moments of inertia at the two considered shape
configurations are rather similar
when the pairing correlations are included,
and are both much smaller than the rigid-body values
even at the largest value of the deformation parameters.
If the pairing correlations are set to zero,
the calculated moment of inertia for the quadrupole shape becomes much larger
and approaches to the rigid-body value
as it is observed from the high-spin limit.
However, the effect of pairing correlation on the inertia
for the tetrahedral shape is rather small,
even though the pairing gap takes more or less the same values
as in the case of the quadrupole shape.
The reason why the moment of inertia is small and
is affected very weakly by the pairing correlation
may be because the chosen nucleus in this case is
the tetrahedral doubly-closed shell nucleus.
The shell gap is $\sim 1.5-2$ MeV and is larger than the pairing gap,
which is in contrast to the case of quadrupole deformation,
where the mean single-particle level spacing is
much smaller than the pairing gap.


\subsection{ Tetrahedral spectra in $^{110}$Zr}
\label{sec:zr110}

The nuclear potential energy surfaces for the doubly-magic Zirconium nuclei
have been studied in Ref.~\cite{Dudek09} and the corresponding illustrations
obtained using the phenomenological approach
with the Woods-Saxon mean-field Hamiltonian
can be found in figure~3 of the above reference.
The symmetry-oriented discussion of the corresponding shell-effects
can be found in Ref.~\cite{KMa09}.
A discussion of the static-energy properties in a few nuclei
in the vicinity of $^{110}$Zr using Hartree-Fock approach
can be found in Ref.~\cite{NSc04},
whereas the tetrahedral rotational properties,
specifically for the nucleus $^{110}$Zr,
have been studied using the cranking-Skyrme-Hartree-Fock method
in Ref.~\cite{NSc06} and using the methods similar to
that of the present article in Ref.~\cite{YKIS2011}.

In the present work the method of calculation is essentially
the same as in \cite{YKIS2011}, except that
the different Woods-Saxon Hamiltonian parameter set is used.
The mean-field parameters and the force strengths used in
the present calculation are given in Table~\ref{tab:Zr}.
The calculated pairing gaps are used because
the experimental even-odd mass differences are not
available for this unstable nucleus.
The numbers of nodes for the Gaussian quadratures are chosen to be
$N_\alpha=N_\gamma=N_\beta=64$ after verifying the stability conditions
for the final result.

\begin{table}[ht]
\begin{center}
\begin{tabular}{ccccccccc}
\hline
$\alpha_{20}$ & $\alpha_{40}$ &
${\mit\Delta}_{\rm n}$ [MeV] & ${\mit\Delta}_{\rm p}$ [MeV] &
$G_{\rm n}$ [MeV] & $G_{\rm p}$ [MeV] &
$\chi$ [MeV$^{-1}$] & $g_0^{\rm n}$ [MeV] & $g_0^{\rm p}$ [MeV] \cr
\hline
0.333\  & $-$0.026\  & 1.129 & 1.113 &
0.1471 & 0.2698 & 4.785$\times10^{-4}$ & 0.1407 & 0.2625 \cr
\hline
\end{tabular}
\end{center}
\caption{
The calculated ground-state mean-field parameters
($\alpha_{20}$, $\alpha_{40}$,
${\mit\Delta}_{\rm n}$, ${\mit\Delta}_{\rm p}$),
and the force strength parameters determined based on them
for $^{110}_{\ 40}$Zr$^{}_{70}$.
The value $g_2^\tau/g_0^\tau=13.6$ is taken for
the ratio of the quadrupole and monopole pairing.
}
\label{tab:Zr}
\end{table}

\begin{figure}[!ht]
\begin{center}
\includegraphics[width=140mm]{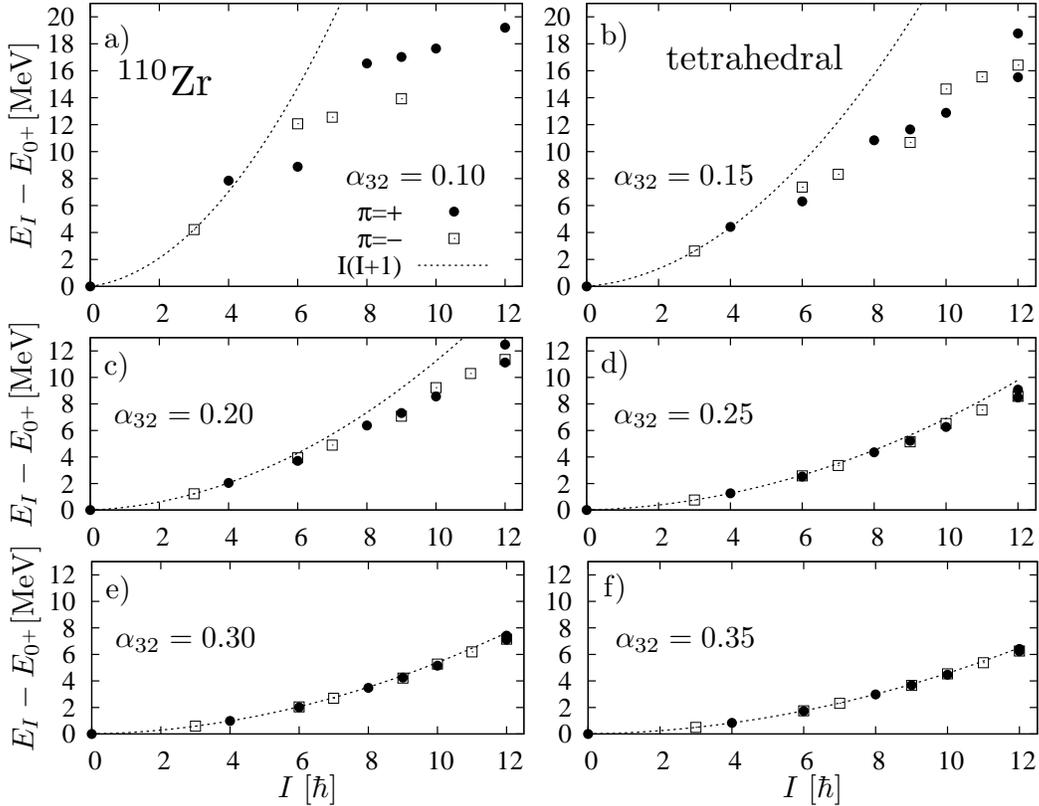}
\vspace*{-4mm}
\caption{
Calculated spectra of tetrahedral states in $^{110}$Zr
with $\alpha_{32}=0.10$, 0.15, 0.20, 0.25, 0.30, and 0.35,
respectively, for a), b), c), d), e), and f).
The figure is similar to
that in Ref.~\cite{YKIS2011} but only the lowest band is selected
and the results for larger deformation are included.
Note that almost exact degeneracies for $I=(6^+,6^-),
(9^+,9^-),(10^+,10^-),(2 \times 12^+,12^-)$ states
are obtained for $\alpha_{32}\ge 0.30$.
}
\label{fig:Zrtall}
\end{center}
\end{figure}

\begin{figure}[!ht]
\begin{center}
\includegraphics[width=140mm]{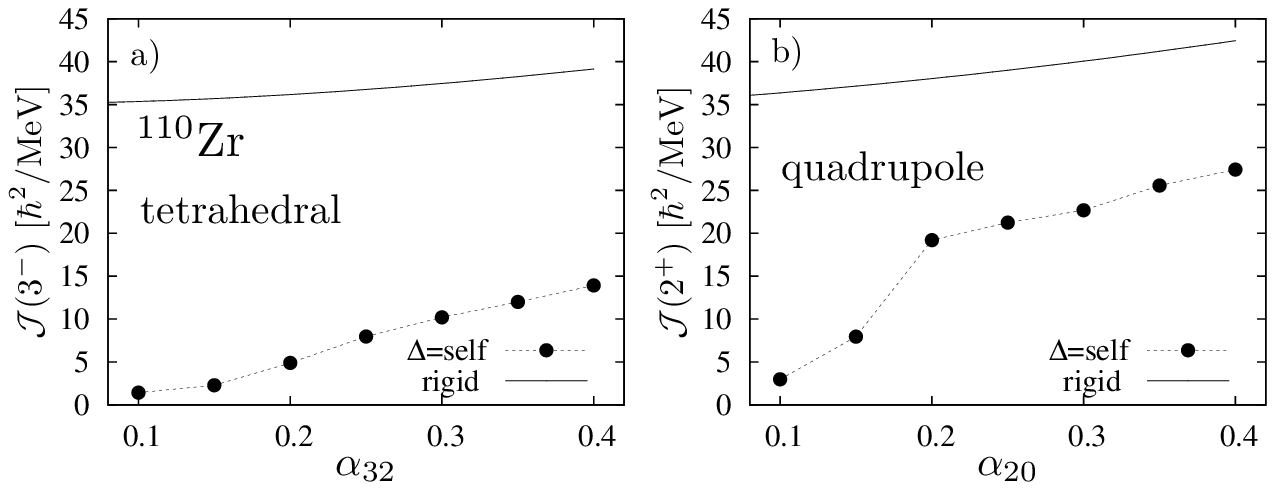}
\vspace*{-4mm}
\caption{
Moment of inertia estimated from the calculated spectra
for the pure tetrahedral states (left) and the pure quadrupole states (right)
in $^{110}$Zr.
The energy of the first excited $3^-$ ($2^+$)
is used for estimation of the former (latter).
The rigid-body moment of inertia are shown as solid lines.
}
\label{fig:Zrmoi}
\end{center}
\end{figure}

In Figure~\ref{fig:Zrtall} we show the excitation energies
for various deformations in function of angular momentum,
obtained using the angular-momentum and parity projection techniques.
The cranking axis is chosen to be the $y$-axis
[$\theta=0^\circ$ in Eqs.~(\ref{eq:cpdefh}) and~(\ref{eq:tiltangle})].
As it is seen from the Figure, the over-all pattern of the excitation scheme
resembles the one in $^{160}$Yb.
However, compared with the results for $^{160}$Yb of Fig.~\ref{fig:Ybtall},
the transition to the ideal rotor occurs slower,
i.e., it occurs at the larger deformation in the lighter system $^{110}$Zr.
In particular, the energy-vs.-spin relation resembles a rigid rotation
only for $\alpha_{32} \gtsim 0.30$.
Compared with the results in Ref.~\cite{YKIS2011},
those in the present work are very similar,
indicating that the choice of the parameter set of the WS potentials
very little affects the rotational properties of
a tetrahedral symmetric nucleus
-- provided a realistic choice of parameters is used.
We have checked the dependence of the projected energies on
the tilting angle of the cranking axis also for the nucleus $^{110}$Zr
(cf. also figure~1 in Ref.~\cite{NSc05}).
Again, the result is found to stay the same,
when the tilting angle is changed
in the same way as in $^{160}$Yb [cf. Eq.~(\ref{eq:tiltangle})].

In Fig.~\ref{fig:Zrmoi}, the moments of inertia for the tetrahedral
and quadrupole deformations are compared for the case of $^{110}$Zr;
only the results with pairing correlations are presented.
As it is seen, the values of the moments of inertia
for two types of deformations are slightly different;
the ratio ${\cal J}(3^-)/{\cal J}_{\rm rigid}$ for the tetrahedral shape
is considerably smaller than the ratio ${\cal J}(2^+)/{\cal J}_{\rm rigid}$
for the quadrupole shape.
One of the reasons may be traced back to the somewhat larger shell gap
at $Z=40$ for the tetrahedral shape, so that
the ratio ${\cal J}(3^-)/{\cal J}_{\rm rigid}$ in $^{110}$Zr
is smaller than that in $^{160}$Yb.
Furthermore, the pairing gaps for the quadrupole shape
with $\alpha_{20} \ge 0.20$ are somewhat reduced in $^{110}$Zr,
which increases the quadrupole moment of inertia;
those combined effects for the two types of shapes may make the difference
of their behaviour in $^{110}$Zr compared to the case of $^{160}$Yb
(and of $^{226}$Th, see below).


\subsection{ Tetrahedral spectra in $^{226}$Th}
\label{sec:th226}

The calculation procedure for $^{226}$Th is the same as that
for $^{160}$Yb and $^{110}$Zr. The mean-field single-nucleon energies
for this particular nucleus in function of tetrahedral deformation
can be found in figure~4 of Ref.~\cite{JDu03}.
The parameters determined by the Woods-Saxon-Strutinsky calculation,
and those of the force strengths are tabulated in Table~\ref{tab:Th}.
They are similar to the ones used in Ref.~\cite{TS12},
where the $\alpha_{30}$ deformation was also taken into account.
Slightly different values of the parameters as compared to those
in Ref.~\cite{TS12} are mainly due to the fact that the different parameter set
of the WS potential is employed.
The computing time of the projection calculation increases
dramatically with increase of the nucleon number
as well as the numbers of the mesh points for Gaussian quadratures.
We have carefully tuned the latter numbers for $^{226}$Th
to obtain the same accuracy as in the case of $^{160}$Yb.  Thus, we take
$N_\alpha=N_\gamma=64$ and $N_\beta=74$ for $\alpha_{32}=0.10-0.15$,
$N_\alpha=N_\gamma=104$ and $N_\beta=84$ for $\alpha_{32}=0.20$,
$N_\alpha=N_\gamma=124$ and $N_\beta=104$ for $\alpha_{32}= 0.25-0.30$,
and $N_\alpha=N_\gamma=N_\beta=124$ for $\alpha_{32}=0.35-0.40$.

\begin{table}[ht]
\begin{center}
\begin{tabular}{ccccccccc}
\hline
$\alpha_{20}$ & $\alpha_{40}$ &
${\mit\Delta}_{\rm n}$ [MeV] & ${\mit\Delta}_{\rm p}$ [MeV] &
$G_{\rm n}$ [MeV] & $G_{\rm p}$ [MeV] &
$\chi$ [MeV$^{-1}$] & $g_0^{\rm n}$ [MeV] & $g_0^{\rm p}$ [MeV] \cr
\hline
0.161\  & 0.093\  & 0.814 & 0.830 &
0.09772 & 0.1289 & 1.744$\times10^{-4}$ & 0.09568 & 0.1267 \cr
\hline
\end{tabular}
\end{center}
\caption{
The calculated ground state deformation parameters
($\alpha_{20}$, $\alpha_{40}$),
the 4-th order even-odd mass difference
$({\mit\Delta}_{\rm n}$, ${\mit\Delta}_{\rm p}$),
and the force strength parameters determined based on them
for $^{226}_{\ 90}$Th$^{}_{136}$.
The value $g_2^\tau/g_0^\tau=13.6$ is taken for
the ratio of the quadrupole and monopole pairing.
}
\label{tab:Th}
\end{table}

\begin{figure}[!ht]
\begin{center}
\includegraphics[width=140mm]{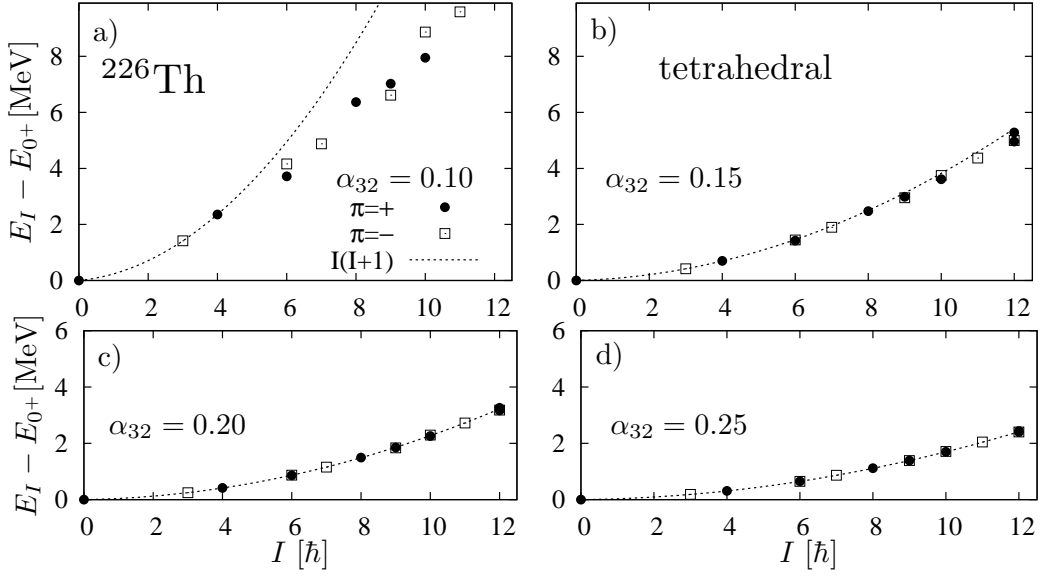}
\vspace*{-4mm}
\caption{
Calculated spectra of tetrahedral-symmetry states in $^{226}$Th
with $\alpha_{32}=0.10$, 0.15, 0.20, and 0.25
respectively, for a), b), c), and d).
Note that almost exact degeneracies for $I=(6^+,6^-),
(9^+,9^-),(10^+,10^-),(2 \times 12^+,12^-)$ states
are obtained for $\alpha_{32}\ge 0.20$.
}
\label{fig:Thtall}
\end{center}
\end{figure}

The result of our angular-momentum and parity projection calculations
for various tetrahedral deformations are shown
in Fig.~\ref{fig:Thtall}.  The characteristic features of the spectra
resemble those of $^{160}$Yb and $^{110}$Zr.
Again, the energy-vs.-spin dependence has approximately linear behaviour
for smaller deformations,
whereas it approaches a parabolic form at increasing deformation.
Comparing the energy-vs.-spin dependence of these three nuclei,
the transition from a linear to parabolic spin dependence occurs
at smaller deformation in nuclei with larger mass number.
More precisely, the energy-vs.-spin dependence becomes almost parabolic
in the following proportions:
at $\alpha_{32}\approx 0.15$ in $^{226}$Th,
at $\alpha_{32}\approx 0.20$ in $^{160}$Yb, and
at $\alpha_{32}\approx 0.25$ in $^{110}$Zr.
This is intuitively acceptable because the concept of the symmetry-breaking
is more and more appropriate for heavier nuclear systems.

\begin{figure}[!ht]
\begin{center}
\includegraphics[width=140mm]{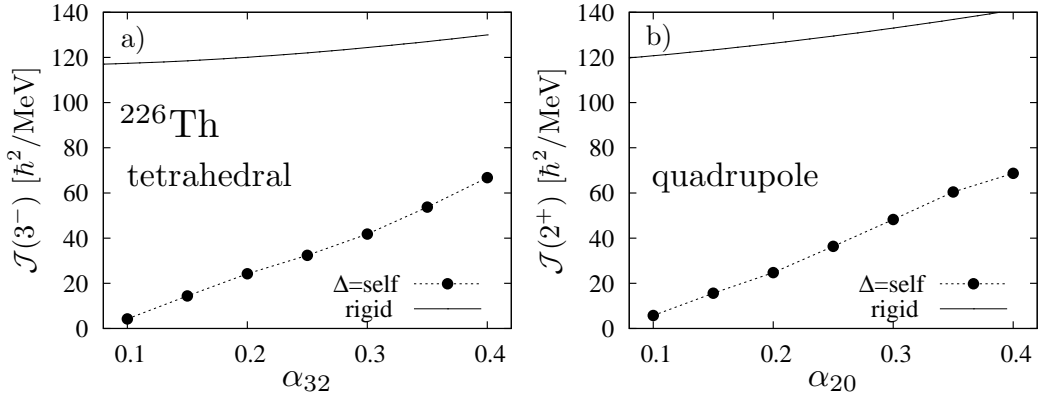}
\vspace*{-4mm}
\caption{
Moments of inertia estimated from the calculated spectra
for the pure tetrahedral states (left) and the pure quadrupole states (right)
in $^{226}$Th.  The energy of the first excited $3^-$ ($2^+$)
is used for estimation of the former (latter).
The classical rigid-body moments of inertia are shown as solid lines.
}
\label{fig:Thmoi}
\end{center}
\end{figure}

The calculated moments of inertia as functions
of quadrupole and tetrahedral deformation parameter are illustrated
in Fig.~\ref{fig:Thmoi},
where only the result including the pairing correlation is shown.
Again, the moments of inertia with the pairing correlation are considerably
smaller than the rigid-body values and they increases with deformation.
The values of moment of inertia are rather similar for the tetrahedral
and quadrupole shapes in $^{226}$Th
as in the case of $^{160}$Yb in Fig.~\ref{fig:Ybmoi}.


\section{Summary}
\label{sec:sum}

We have studied the rotational nuclear properties for the pure tetrahedral
deformation by the angular-momentum and parity projection method
employing the realistic Woods-Saxon mean-field potential
and the schematic separable two-body interaction consistent with it.
In this work we have chosen the tetrahedral
doubly-closed shell nuclei $^{160}$Yb ($Z=70$ and $N=90$),
$^{110}$Zr ($Z=40$ and $N=70$) (see also Ref.~\cite{YKIS2011} for this nucleus)
and $^{226}$Th ($Z=90$ and $N=136$) as illustrative examples.
We have found out that the characteristic spectra
for the totally symmetric representation, i.e.
the $A_1$ irreducible representation of the tetrahedral group $T_d$,
are obtained with the specific sequence composed of
$\ 0^+$, $3^-$, $4^+$, $6^+$, $6^-$, $7^-$, $8^+$,$\cdots$.
These spectra are consistent with those of the simple tetrahedral rotor,
although the spectra are more vibrational-like for small deformations.
However, it is important to emphasise that only
the specific spin-parity combinations appear in the projection calculations
in this work in agreement with the group-theory predictions,
approaching the rotational pattern closer and closer
with increasing tetrahedral deformation.

The Coriolis $|\Delta K|=1$ mixing is introduced
in order to break the time-reversal invariance and
to obtain more reliable estimate of moment of inertia;
only the slope of energy-vs.-spin relation changes
and the qualitative features are not affected by this mixing.
It has been also checked that the results of projected spectra are
independent of the tilting angle of the cranking axis,
which is consistent with the picture that the tetrahedral rotor possesses
a `spherically-symmetric tensor of inertia'.

Stable rigid-rotor energy dependence [$\propto I(I+1)$] appears
for larger deformations in the range studied in this article, but
an approximately linear dependence is obtained with decreasing deformations.
This transition between the vibrational-like and the rotational pattern
occurs at smaller deformation for nuclei with larger mass number.
The moment of inertia for the tetrahedral shape increases
with increasing deformation, but it is much smaller than the rigid-body value
irrespective of the pairing correlations.
The impact of the pairing seems rather limited for tetrahedral deformed nuclei
at least for the doubly-closed shell nuclei,
with relatively large single-particle energy-gaps.

In this work, we have concentrated on the lowest-energy rotational sequences
and shown by the calculations that their properties resemble
the ones characteristic for the totally symmetric,
the so-called $A_1$ irreducible representation of the tetrahedral point-group,
$T_d$, believed to be characteristic for the lowest-energy sequence of states
in fully paired even-even nuclei.
From the group theoretical consideration, it is expected that
the other irreducible representations, i.e., $A_2$, $E$ and/or $F_1$, $F_2$,
would appear in the excited rotational bands of even-even nuclei,
see the results of Ref.~\cite{YKIS2011},
or in the spectra for odd-odd nuclei.
Moreover, there exist different types of representations
associated with the so-called double tetrahedral group,
specific for the odd nuclei, see Appendix.

We believe that both the experimental analysis as well as
the theoretical calculations of the discussed spectral properties
will need to include more levels in the future.
For this purpose the remaining irreducible representations
of the symmetry group may need to be studied compared to
the scalar representations that we focused on in this article.
The analysis of the irreducible representation structure of the solutions
is important for the next step of the analysis,
which would consist in calculating the electromagnetic transitions
and the branching ratios, the observables which change rapidly
with symmetry of the system. This step of the analysis will be essential
for establishing the experimental criteria of determining the presence
of tetrahedral symmetry in the physics of subatomic systems.


\section*{ACKNOWLEDGEMENTS}

We appreciate fruitful discussions with Masayuki Matsuo especially
on the content of Appendix.
This work is supported in part by Grant-in-Aid for Scientific Research (C)
No.~22540285 from Japan Society for the Promotion of Science.


\appendix*
\section{Spin-parity relations in a tetrahedrally symmetric rotor}

Although it may be considered a textbook matter,
some spin-parity properties of the rotational energies of
the tetrahedrally symmetric rotor will be summarised in this Appendix,
to facilitate the comparison between the results of
the microscopic calculations with the projection techniques
as obtained in this article and the group-theory expectations
(see e.g. Ref.~\cite{Hamermesh}).
The representation of the rotor states
with definite spin-parity $I\pi$ ($\pi=\pm$), $D^{(I\pi)}$,
which have a certain symmetry governed by a group $G$,
can be decomposed into its irreducible representations, $D_i$ ($i=1,\cdots,M$),
with multiplicity $a_i^{(I\pi)}$;
\begin{equation}
 D^{(I\pi)}=\sum_{i=1}^M  a_i^{(I\pi)} D_i.
\label{eq:decompo}
\end{equation}
The multiplicity can be calculated by the standard formula~\cite{Hamermesh},
\begin{equation}
 a_i^{(I\pi)}=\frac{1}{N_G}\sum_{R\in G}\chi_{I\pi}(R)\chi_i(R)
 =\frac{1}{N_G}\sum_{\alpha=1}^M g_\alpha\chi_{I\pi}(R_\alpha)\chi_i(R_\alpha),
\label{eq:Charn}
\end{equation}
where the number $N_G$ is the order of the group $G$,
$\chi_{I\pi}(R)$ and $\chi_i(R)$ are the characters of
the representations $D^{(I\pi)}$ and $D_i$, respectively,
for the group element $R$,
and the quantity $g_\alpha$ denotes the number of elements
in the class $\alpha$, whose representative element is $R_\alpha$.
Note that the decomposition~(\ref{eq:decompo}) is performed by
a unitary transformation in the $(2I+1)$ dimensional space of
the rotor wave functions for given $I\pi$;
more precisely, a specific combination of the $K$-mixing
generates each irreducible representation.

\begin{table}[ht]
\begin{center}
\begin{tabular}{c|rrrrr}
$T_d$ & $E$\ \ & $\ C_3(8)$ & $\ C_2(3)$ & $\ \sigma_d(6)$ & $\ S_4(6)$ \cr
\hline
$A_1$ & \ \ 1\ \  & 1\ \  & 1\ \  & 1\ \  & 1\ \  \cr
$A_2$ & \ \ 1\ \  & 1\ \  & 1\ \  & $-1$\ \  & $-1$\ \  \cr
$E$   & \ \ 2\ \  & $-1$\ \  & 2\ \  & 0\ \  & 0\ \  \cr
$F_1(T_1)$ & \ \ 3\ \  & 0\ \  & $-1$\ \  & $-1$\ \ & 1\ \  \cr
$F_2(T_2)$ & \ \ 3\ \  & 0\ \  & $-1$\ \  & 1\ \  & $-1$\ \  \cr
\end{tabular}
\end{center}
\caption{
Character table for the tetrahedral group $T_d$.
Taken from Ref.~\cite{Hamermesh}
(note that $C_2=S_4^2$ and $F_{1,2}$ are sometimes denoted as $T_{1,2}$).
}
\label{tab:charTd}
\end{table}

\begin{table}[ht]
\begin{center}
\begin{tabular}{c|ccccccccccccccccc}
$I^+$
  & \ $0^+$ & $1^+$ & $2^+$ & $3^+$ & $4^+$ & $5^+$ & $6^+$ & $7^+$ & $8^+$ &
    $9^+$ & $10^+$ & $11^+$ & $12^+$ & $13^+$ & $14^+$ & $15^+$ & $16^+$ \cr
\hline
$A_1$ & 1 & 0 & 0 & 0 & 1 & 0 & 1 & 0 & 1 & 1 & 1 & 0 & 2 & 1 & 1 & 1 & 2 \cr
$A_2$ & 0 & 0 & 0 & 1 & 0 & 0 & 1 & 1 & 0 & 1 & 1 & 1 & 1 & 1 & 1 & 2 & 1 \cr
$E$   & 0 & 0 & 1 & 0 & 1 & 1 & 1 & 1 & 2 & 1 & 2 & 2 & 2 & 2 & 3 & 2 & 3 \cr
$F_1(T_1)$
      & 0 & 1 & 0 & 1 & 1 & 2 & 1 & 2 & 2 & 3 & 2 & 3 & 3 & 4 & 3 & 4 & 4 \cr
$F_2(T_2)$
      & 0 & 0 & 1 & 1 & 1 & 1 & 2 & 2 & 2 & 2 & 3 & 3 & 3 & 3 & 4 & 4 & 4 \cr
\end{tabular}

\vspace*{5mm}

\begin{tabular}{c|ccccccccccccccccc}
$I^-$
  & \ $0^-$ & $1^-$ & $2^-$ & $3^-$ & $4^-$ & $5^-$ & $6^-$ & $7^-$ & $8^-$ &
    $9^-$ & $10^-$ & $11^-$ & $12^-$ & $13^-$ & $14^-$ & $15^-$ & $16^-$ \cr
\hline
$A_1$ & 0 & 0 & 0 & 1 & 0 & 0 & 1 & 1 & 0 & 1 & 1 & 1 & 1 & 1 & 1 & 2 & 1 \cr
$A_2$ & 1 & 0 & 0 & 0 & 1 & 0 & 1 & 0 & 1 & 1 & 1 & 0 & 2 & 1 & 1 & 1 & 2 \cr
$E$   & 0 & 0 & 1 & 0 & 1 & 1 & 1 & 1 & 2 & 1 & 2 & 2 & 2 & 2 & 3 & 2 & 3 \cr
$F_1(T_1)$
      & 0 & 0 & 1 & 1 & 1 & 1 & 2 & 2 & 2 & 2 & 3 & 3 & 3 & 3 & 4 & 4 & 4 \cr
$F_2(T_2)$
      & 0 & 1 & 0 & 1 & 1 & 2 & 1 & 2 & 2 & 3 & 2 & 3 & 3 & 4 & 3 & 4 & 4 \cr
\end{tabular}
\end{center}
\caption{
The number of states $a_i^{(I\pi)}$
belonging to the five irreducible representations of $T_d$
for integer spins; those for each parity are separately shown.
}
\label{tab:spTd}
\end{table}

The $T_d$ group has five irreducible representations and classes,
whose representative elements are
$E$, $C_2\,(=S_4^2)$, $C_3$, $\sigma_d$, and $S_4$;
see Ref.~\cite{Hamermesh} for the notation.
The characters for each irreducible representation
are listed in Table~\ref{tab:charTd} for completeness, and
those for the rotor representation~\cite{Wil35} are as follows;
\begin{equation}
 \chi_{I\pi}(E)=2I+1,\quad
 \chi_{I\pi}(C_n)=\sum_{K=-I}^{I} e^{\frac{2\pi K}{n}i}
 =\frac{\sin{\frac{(2I+1)\pi}{n}}}{\sin{\frac{\pi}{n}}},
\label{eq:charRot}
\end{equation}
\begin{equation}
 \chi_{I\pi}(\sigma_d)=\pi\times\chi_{I\pi}(C_2),\quad
 \chi_{I\pi}(S_4)=\pi \times\chi_{I\pi}(C_4).
\label{eq:charRota}
\end{equation}
Combination of the characters in Eqs.~(\ref{eq:charRot})$-$(\ref{eq:charRota})
and in Table~\ref{tab:charTd} with the formula~(\ref{eq:Charn})
leads to the multiplicities, $a_i^{(I\pi)}$,
which are summarised in Table~\ref{tab:spTd} for integer spins up to $I=16$.
It is easy to verify that
$a_{A_1}^{(I\pm)}= a_{A_2}^{(I\mp)}$,
$a_{E}^{(I+)}= a_{E}^{(I-)}$, and
$a_{F_1}^{(I\pm)}= a_{F_2}^{(I\mp)}$.
In the table, $a_i^{(I\pi)}=0$ means that such states are not allowed,
and $a_i^{(I\pi)}=2$ means that states are doubly degenerate.
In this way the characteristic spin-parity
for the $A_1$ representation in Eq.~(\ref{eq:TdA1}) follows.

\begin{table}[ht]
\begin{center}
\begin{tabular}{c|rrrrrrrr}
$T^D_d$ & \multicolumn{2}{c}{\ $E$} & \multicolumn{2}{c}{\ \ $C_3(8)$} &
 \ $C_2(3)$ & \ $\sigma_d(6)$ & \multicolumn{2}{c}{\ \ $S_4(6)$} \cr
\hline
$E_{1/2}(E'_1)$ & \ 2\ &\ $-2$\  &\ 1\ & $-1$\ &0\ \ \ \ &0\ \ \ \ &
 $\sqrt{2}$\ & $-\sqrt{2}$\  \cr
$E_{5/2}(E'_2)$ & \ 2\ &\ $-2$\  &\ 1\ & $-1$\ &0\ \ \ \ &0\ \ \ \ &
 $-\sqrt{2}$\ & $\sqrt{2}$\  \cr
$G_{3/2}(G')$ & \ 4\ &\ $-4$\  &\ $-1$\ & 1\ &0\ \ \ \ &0\ \ \ \ &
 \ 0 & 0\ \ \cr
\end{tabular}
\end{center}
\caption{
Character table specific for the extended (also called `double')
tetrahedral group $T^D_d$.
The second entry in the corresponding columns denotes the characters
of extended elements.
[Taken from Ref.~\cite{HerzbIII}
(note that $E_{1/2}$, $E_{5/2}$ and $G_{3/2}$ are sometimes denoted
as $E'_1$, $E'_2$ and $G'$).]
}
\label{tab:charTd1}
\end{table}

\begin{table}[ht]
\begin{center}
\begin{tabular}{c|cccccccccccccccc}
$I^+$
 &\ $\frac{1}{2}^+$ & $\frac{3}{2}^+$ & $\frac{5}{2}^+$ & $\frac{7}{2}^+$ &
 $\frac{9}{2}^+$ & $\frac{11}{2}^+$ & $\frac{13}{2}^+$ & $\frac{15}{2}^+$ &
 $\frac{17}{2}^+$ & $\frac{19}{2}^+$ & $\frac{21}{2}^+$ & $\frac{23}{2}^+$ &
 $\frac{25}{2}^+$ & $\frac{27}{2}^+$ & $\frac{29}{2}^+$ & $\frac{31}{2}^+$ \cr
\hline
$E_{1/2}(E'_1)$&
 \ 1 & 0 & 0 & 1 & 1 & 1 & 1 & 1 & 2 & 2 & 1 & 2 & 3 & 2 & 2 & 3 \cr
$E_{5/2}(E'_2)$&
 \ 0 & 0 & 1 & 1 & 0 & 1 & 2 & 1 & 1 & 2 & 2 & 2 & 2 & 2 & 3 & 3 \cr
$G_{3/2}(G')$&
 \ 0 & 1 & 1 & 1 & 2 & 2 & 2 & 3 & 3 & 3 & 4 & 4 & 4 & 5 & 5 & 5 \cr
\end{tabular}

\vspace*{5mm}

\begin{tabular}{c|cccccccccccccccc}
$I^-$
 &\ $\frac{1}{2}^-$ & $\frac{3}{2}^-$ & $\frac{5}{2}^-$ & $\frac{7}{2}^-$ &
 $\frac{9}{2}^-$ & $\frac{11}{2}^-$ & $\frac{13}{2}^-$ & $\frac{15}{2}^-$ &
 $\frac{17}{2}^-$ & $\frac{19}{2}^-$ & $\frac{21}{2}^-$ & $\frac{23}{2}^-$ &
 $\frac{25}{2}^-$ & $\frac{27}{2}^-$ & $\frac{29}{2}^-$ & $\frac{31}{2}^-$ \cr
\hline
$E_{1/2}(E'_1)$&
 \ 0 & 0 & 1 & 1 & 0 & 1 & 2 & 1 & 1 & 2 & 2 & 2 & 2 & 2 & 3 & 3 \cr
$E_{5/2}(E'_2)$&
 \ 1 & 0 & 0 & 1 & 1 & 1 & 1 & 1 & 2 & 2 & 1 & 2 & 3 & 2 & 2 & 3 \cr
$G_{3/2}(G')$&
 \ 0 & 1 & 1 & 1 & 2 & 2 & 2 & 3 & 3 & 3 & 4 & 4 & 4 & 5 & 5 & 5 \cr

\end{tabular}
\end{center}
\caption{
The number of states $a_i^{(I\pi)}$
belonging to the three irreducible representations specific for $T^D_d$
for half-odd integer spins; those for each parity are separately shown.
}
\label{tab:spTd1}
\end{table}

As for the excitations of odd nuclei, i.e., for half-odd integer spins $I$,
the same calculation can be done,
but one has to consider the extended rotation group (the double group) $G^D$,
see e.g. Ref.~\cite{HerzbIII}
(or equivalently, the two-valued representations~\cite{Hamermesh}).
In the extended group, the number of elements is doubled by
extending the range of rotational angle about an axis from $2\pi$ to $4\pi$,
because the $2\pi$ rotation is not the identity operation
but changes sign for the rotor states with half-integer spins.
The character table specific for the double tetrahedral group $T^D_d$
is shown in Table~\ref{tab:charTd1},
and the resultant multiplicities are given in Table~\ref{tab:spTd1},
where $a_{E_{1/2}}^{(I\pm)}= a_{E_{5/2}}^{(I\mp)}$
and $a_{G_{3/2}}^{(I+)}= a_{G_{3/2}}^{(I-)}$ can be easily confirmed.
Table~\ref{tab:spTd1} can be also used to see how a spherical
single-particle orbit $j^\pi$ decomposes into
the two-fold ($E_{1/2,\,5/2}$) and four-fold ($G_{3/2}$)
degenerate orbits for finite tetrahedral deformation.


\end{document}